# Van der Waals heterostructure metasurfaces: atomic-layer assembly of ultrathin optical cavities

Luca Sortino[1*], Jonas Biechteler[1], Lucas Lafeta[2], Lucca Kühner[1], Achim Hartschuh[2], Leonardo de S. Menezes[1,3], Stefan A. Maier[4,5], Andreas Tittl[1*]

[1]Chair in Hybrid Nanosystems, Nanoinstitute Munich, Faculty of Physics, Ludwig-Maximilians-Universität München, 80539 Munich, Germany

[2]Department of Chemistry, Ludwig-Maximilians-Universität München, Butenandtstraße. 5-13, 81377, Munich, Germany

[3]Departamento de Física, Universidade Federal de Pernambuco, 50670-901 Recife-PE, Brazil

[4]School of Physics and Astronomy, Monash University, Wellington Rd, Clayton VIC 3800, Australia

[5]The Blackett Laboratory, Department of Physics, Imperial College London, London, SW7 2AZ, United Kingdom

*Corresponding author. Email: luca.sortino@physik.uni-muenchen.de, Andreas.Tittl@physik-uni.muenchen.de

**Photonics has been revolutionized by breakthroughs in optical metasurfaces and layered two-dimensional materials. Yet, integrating these two fields in a singular system has remained challenging. Here, we introduce the concept of van der Waals (vdW) heterostructure metasurfaces, where ultrathin multilayer vdW material stacks are shaped into precisely engineered resonant nanostructures for boosting light-matter interactions. By leveraging quasi-bound states in the continuum to create intrinsic cavities from $WS_2$ monolayers encapsulated in hexagonal boron nitride, we observe room-temperature strong coupling and polaritonic luminescence, which further unveils a saturation of the strong-coupling regime at ultralow fluences <1 nJ/cm$^2$, more than three orders of magnitude smaller than in previous 2D-cavity systems. Our approach, seamlessly merging metasurfaces and vdW materials, unlocks new avenues for ultrathin optical devices with atomic-scale precision and control.**

Metasurfaces are periodic two-dimensional (2D) arrays of optically resonant nanostructures that have revolutionized the field of optics by enabling flat optical components with a vast design parameter space (*1, 2*) and an ever-growing set of applications, from metalenses to polarization optics (*3*), mainly stemming from the engineering of the complex amplitude and phase of light interacting with each single resonant nanostructure (*4*). Another aspect, from plasmonic and nanophotonic concepts, is the engineering of electromagnetic resonances in nanostructured arrays to achieve optical resonances with high quality (Q) factors (*5, 6*), defined as the ratio of the resonance frequency to its linewidth, crucial for applications in sensing (*7, 8*) and nonlinear phenomena (*9, 10*). Due to the efficient trapping of resonant photons, strong electromagnetic fields associated with high Q factor resonances placed optical cavities at the forefront of research in quantum electrodynamics in semiconductors since the early 1990s (*11*). However, in the field of optical metasurfaces, active layers are integrated with metasurfaces as two distinct systems (*12*), only coupling to the weak evanescent fields outside the resonant structures (*13*).

The field of 2D materials, started with graphene more than twenty years ago, has revolutionized optoelectronics and the level of control in the deterministic assembly of nanostructures down to single atomic layers (*14*). More broadly, van der Waals (vdW) materials refer to a family of crystals where atoms are covalently bonded along the in-plane direction, while in the out-of-plane direction only weak vdW forces hold the crystalline planes together. Unlike conventional materials, this structure enables the mechanical exfoliation of crystal planes down to single atomic layers. The deterministic manipulation of atomic layers opens up ultimate control over the composition of artificial structures, with the possibility of designing active materials with tailored optoelectronic properties. A key technological advancement is the capability of vdW materials for vertically stacking arbitrary 2D layers, enabled by their unconventional crystal structure. Contrary to conventional semiconductor crystals, vdW materials lack dangling bonds and lattice mismatch. As such, they can be deterministically stacked to form vertical vdW heterostructures, composed of two or more different 2D materials, selected from the vast catalog of vdW crystals. Among them, hexagonal boron nitride (hBN), a wide bandgap dielectric, has found widespread use as a transparent substrate and encapsulation material for the fabrication of high quality vdW heterostructures (*15, 16*). In the case of 2D semiconductors, transition metal dichalcogenide (TMDCs) monolayers provide exceptional properties for light-matter coupling, from stable excitons at room temperature (*17, 18*), to valley physics and single-photon sources (*19*), with additional degrees of freedom when realizing heterostructures with two or more TMDC layers. hBN encapsulation has now become a standard process in the fabrication of vdW heterostructures, for preserving and improving the optical and electronic qualities of the fabricated devices (*20*) which are driving significant progress in condensed matter physics, enabling discoveries on Moiré effects (*21, 22*), topological states (*23*), 2D magnetism (*24*), all in sub-micron thick devices.

Beyond their atomically thin form, the bulk counterparts of vdW materials (50-500 nm thickness) likewise garnered significant interest in nanophotonics due to their high refractive indexes (*25*) and large anisotropies (*26*). Nanostructuring vdW materials for engineering optical resonances, as in conventional semiconductor-on-insulator architectures, emerged as a promising avenue in nanophotonics, providing a novel approach to designing ultrathin photonic devices (*27, 28*). In addition to the encapsulation properties mentioned above, hBN has also shown exceptional promise as a nanophotonic platform, from sub-diffractional optical cavities (*29, 30*) to nanophotonic circuitry (*31*), with the additional presence of single photon sources and phonon-polaritons (*32*). In the context of metasurface design, hBN is a versatile material; by leveraging quasi-bound states in the continuum (qBIC) physics (*33, 34*), hBN metasurfaces sustaining high Q resonances (Q > 200) across the visible have been realized (*35*). Such qBIC modes act as resonant optical cavities, strongly confining intense electromagnetic fields in subwavelength thicknesses, and have been demonstrated for the efficient enhancement of light-matter interactions (*36*, *37*) and achieving the strong light-matter coupling regime with excitons in both monolayer and bulk TMDCs (*38*, *39*). Thus, the integration of rationally designed atomically thin vdW heterostructures with the photonic degrees of freedom provided by qBIC metasurfaces offers a promising strategy for developing ultrathin photonic devices. The robust light confinement achieved through qBICs can further address the limitations associated with traditional optical cavities, including bulk and costly Bragg mirrors (*40*), or the losses and integration challenges posed by plasmonic particle-on-mirror cavities (*41*).

In this work, we introduce the concept of van der Waals heterostructure metasurfaces (vdW-HM) as illustrated in Fig.1A. By nanopatterning optical metasurface designs within stacked vdW materials, the thin hBN encapsulating layers act as a dielectric metasurface nanocavity (*35*) where qBIC resonances are generated by introducing an asymmetry in the unit cell design (*6*). Notably, this monolithic approach eliminates the need of bulky external cavity systems and provides tunable high-Q resonances within the same ultrathin heterostructure. We fabricated vdW-HM from monolayer $WS_2$ encapsulated in hBN, as a prototypical vdW heterostructure, and demonstrated the strong light-matter coupling regime between qBIC modes and 2D excitons, observing Rabi splitting energy above 30 meV at room temperature. We characterize the metasurface-coupled exciton-polaritons via reflectance and photoluminescence (PL) emission with angle-resolved spectroscopy, and report strong nonlinearities under increasing excitation power, leading to a saturation of the strong coupling regime at fluences <1 nJ/cm$^2$, a factor of more than $10^3$ smaller than that observed for TMDCs in standard optical cavities (*42*). These results are promising for ultrafast switches, lasing and bosonic condensates in nanoscale photonic structures (*43*). The rich physics and optical properties of 2D materials, such as spin-valley physics, single photon sources and multibody exciton complexes, combined with the ultimate control over their constituents down to single atomic layers, pave the way for advanced ultrathin photonics based on optical metasurfaces.

**Principles of light-matter coupling in vdW heterostructure metasurfaces** The enhancement of light-matter interaction with 2D semiconductors in optical metasurfaces requires strongly localized in-plane electromagnetic fields, necessary to efficiently couple the optical mode with 2D excitons confined in the atomically thin monolayer (Fig. 1B). Dielectric metasurfaces supporting photonic bound states in the continuum (*33*, *34*) excel at controlling light-matter interactions at sub-micron thicknesses via engineering radiation losses. By introducing a symmetry breaking in the geometry of the periodically arranged unit, qBIC resonances with Fano-like shapes can be designed to optimize both the spectral position and the linewidth, while providing strongly confined in-plane electromagnetic fields (*36*). The control over the radiative properties of the qBIC mode is achieved by introducing an asymmetry parameter ($\alpha$) in the unit cell design (*6*). A true BIC is a symmetry protected state with a theoretically infinite Q at $\alpha = 0$. When breaking the symmetry ($\alpha > 0$), a radiative channel is opened, allowing the far-field excitation of a qBIC resonance, which can reach extremely high Q factors >$10^7$ (*34*). By modifying $\alpha$, the linewidth of the qBIC resonances can be precisely tailored, for instance, to match the system's critical coupling condition (*38*). Moreover, qBIC resonances can be spectrally shifted throughout broad spectral ranges by simply scaling the unit cell size via a lateral scaling factor (*S*) acting on all the in-plane dimensions of the unit cell design, except for the height, which is usually limited by the active material layer thickness.

We designed, via numerical simulations, hBN-based heterostructure metasurfaces with qBIC modes covering the entire emission spectrum of monolayers $WS_2$ (Fig.1C). Notably, we can selectively adjust the qBIC resonance relative to the exciton spectral position via the scaling factor, $S$, which is linked to the detuning parameter $\delta = E_{qBIC} - E_X$, where $E_{qBIC}$ and $E_X$ are the qBIC mode and WS$_2$ exciton energies, respectively. The qBIC-exciton system can be described analytically with a coupled oscillator model (see Methods for the complete model). When reaching the strong light-matter coupling regime, the coupled exciton-photon system hybridizes with the formation of the lower and upper exciton-polariton branches (Fig.1D). Polaritons inherit the potential for strong coupling from their excitonic component but offer wide tunability given by their photonic functionality and, in semiconductors, they have been used for low-threshold lasers and Bose-Einstein condensates with applications in low power optoelectronics and computation (*44*, *45*).

As the qBIC resonance is tuned across the exciton energy, an anticrossing between the two resonances is formed. At zero detuning, $\delta = 0$, we determine the Rabi splitting energy ($\Omega_R$), which is directly proportional to the coupling strength ($g \sim \mu \cdot E$), defined as the product of the electromagnetic fields ($E$) and transition dipole moment ($\mu$) of a resonant electronic transition. Moreover, Fig.1E further illustrates the angular dispersion of the qBIC-exciton system as a function of the in-plane wavevector $k_x$ (oriented along the long axis of the nanorods), modeled as described in the Methods. The qBIC cavity mode exhibits a negative dispersion (see also Fig.S1), contrary to common optical cavities, which has been linked to a favorable configuration for achieving condensation in saddle-like points (*46*). When at non-zero detuning, the Rabi splitting is shifted to higher angles, where the two resonances cross.

**Experimental realization of vdW heterostructure metasurfaces** The fabrication of vdW metasurfaces begins with the assembly of a vdW heterostructure via mechanical exfoliation and deterministic transfer for each layer. Fig.2A shows an optical image of the fabricated vdW stack, where a WS$_2$ monolayer is encapsulated between two bulk hBN thin films of approximately 60-65 nm in height. We then proceeded with an electron beam lithography writing step, hard mask deposition and reactive ion etching to transfer the metasurface design into the vdW heterostructure (see Methods and Fig.S2 for fabrication details). More than 15 individual metasurface fields with different geometrical parameters were realized within the initial vdW heterostructure area of approximately 120 μm x 70 μm (Fig.2A). The chosen vdW-HM design with a total height of approximately 125 nm is composed of asymmetric hBN/WS$_2$/hBN rod pairs, where $L_0$ is the base rod length and $\Delta L$ gives the asymmetry factor for the generation of the radiative qBIC channel (Fig.2B).

The optical quality of the WS$_2$ monolayer is preserved during the fabrication of the vdW-HM, with negligible differences observed in the PL emission spectra of final samples compared to a reference of unstructured hBN-encapsulated WS$_2$ (Fig.2C). Transmittance measurements from the reference HS and a vdW-HM with $\delta = 15$ meV reveal a distinct splitting in the transmission spectrum at the exciton energy, providing evidence for a strong interaction between 2D excitons and qBIC modes (Fig.2D). We further investigated the coupled exciton-metasurface system, in both optical transmission and PL emission, for the whole set of fabricated metasurfaces. We extracted Q factors above 100 for detuned metasurfaces (Fig.S3) and observed a decrease in Q values at the exciton resonance due to the broadening and splitting of the cavity mode. In Fig.2E, we plot the derivative of the transmittance as a function of the scaling factor, where the detuning of the qBIC cavity and the exciton resonance exhibits clear anti-crossing at the corresponding WS$_2$ exciton energy (Fig.2E, inset), confirming the strong light-matter coupling regime in the ultrathin vdW-HM. When probing the PL emission from the same set of vdW-HM (Fig.2F-G), we observe the presence of a PL peak at lower energies than the WS$_2$ excitons, ascribed to the lower polariton branch. From both datasets, we extracted a Rabi splitting energy of $\Omega_R = 30 \pm 1$ meV at room temperature, following a coupled harmonic oscillator model (see Methods), which is comparable to values found in TMDC monolayers in commonly used optical cavities (*43, 47*). Furthermore, we show in Fig.S4 that the observation of the strong-coupling regime is replicated in a second sample.

**Angle-resolved spectroscopy of hBN/WS$_2$/hBN vdW heterostructure metasurfaces** To fully characterize the exciton-qBIC interactions in detuned metasurfaces, where the Rabi splitting appears at non-zero angles (Fig.1D), we carried out numerical simulations of the metasurface angular dispersion and experimental characterization via back focal plane spectroscopy (*48*) (see Methods and Fig.S5 for experimental details and setup).

Because the periodic qBIC metasurface design essentially constitutes a 2D diffraction grating, we first studied the effect of the diffraction modes on the angular emission. We consider the grating equation $k_q - k_i = \pm m K$, where $k_q$ and $k_i$ are respectively the wavevectors of the diffracted and incident wave, $m$ is an integer number, and K is related to the reciprocal 2D lattice set by the metasurface geometry. Specifically, we carried out rigorous coupled-wave analysis (RCWA) simulations of the angle-resolved reflectance of vdW-HMs both on a glass substrate and in a homogeneous vacuum environment (Fig.3A). In the substrate case, we observed a prominent effect of the diffraction orders in the angular reflectance spectrum (see also Fig.S6), especially noticeable at low angles (Fig.3A, left). In a homogeneous environment, where diffraction orders are absent at low angles due to the lack of substrate effects, the angular dispersion exhibits the predicted anti-crossing behavior (Fig.3A, right).

To validate these effects experimentally, we collected the angle-resolved reflectance of the strongly coupled vdW-HM via back focal plane imaging. Fig.3B shows the reflectance of a positively detuned vdW-HM ($\delta$ = 96 meV), where strong diffraction modes are observed. We observe a splitting of approximately 44 meV at the exciton energy from the corresponding reflectance profile (Fig.3C). Compared with the numerical simulations in Fig.3D (see also Fig.S7) we have an excellent agreement between experimental and numerical results. We then analyzed the PL emission by exciting the sample from the substrate side and collecting the emitted PL signal and relative back focal plane image (see Methods). We collected the PL emission from vdW-HMs with different detunings, as shown in Fig.3E. The PL closely follows the expected angle-resolved dispersion of the strongly coupled system. For positive detuning, the PL emission seems to be mainly redirected into the grating mode. However, in negatively detuned cavities, we observe that the overall weight of the PL emission shifts to the maximum of the qBIC-coupled lower polariton branch at zero angle. This confirms that the additional PL peak observed in the vdW-HMs is not related to a directional emission into the diffraction orders, but the effective polariton emission following the negative dispersion of the qBIC cavity mode. Moreover, we note that this effect is accompanied by an overall enhancement of the collected PL intensity, as shown in Fig.S8, further demonstrating an increased radiative efficiency.

**Nonlinear metasurface-coupled exciton-polaritons** Strong polariton interactions are fundamental towards the realization of lasing and Bose-Einstein condensates (*49*), requiring high nonlinearities or large densities. To further characterize the nonlinear nature of the exciton-polariton states in our vdW-HM, we investigated the PL as a function of the excitation source intensity. Fig.4A shows the angle-resolved PL emission of negatively detuned vdW metasurface, both in the low and high-power regimes. At high fluences, a clear shift of the PL maximum at zero angles to higher energies is observed, together with the absence of the excitonic peak. As the polaritons density is increased via the optical pump, their mutual dipole-dipole interaction induces a nonlinear response, shifting the emission peak to higher energies (*50*), which can lead to population inversion for lasing or, in case of stronger interaction, condensation.

We first investigated the optical transmission of a vdW-HM with small detuning ($\delta$ = 20 meV), exhibiting a splitting in the transmission spectrum at the WS$_2$ exciton energy. The system is excited with a low repetition rate, supercontinuum pulsed laser (see Methods for experimental details). Remarkably, under increasing fluences, we start to observe a saturation of the strong coupling regime and, above $10^{-9}$ J cm$^{-2}$, the collapse of the two absorption peaks into a single, weakly coupled resonance (Fig.4B). The use of a low repetition rate of 2.05 kHz ensures negligible thermal effects underlying the saturation behavior or the presence of long-lived dark states in our system. This transition is thus ascribed to the strong interactions via the increased polariton density, which is quickly saturated by the lack of excited states available for the optical transition and the lowering of the oscillator strength. These observations are further confirmed in the PL emission (Fig.4C). Analogous to the absorption behavior, an emission doublet is observed at low densities, which quickly converges to a single emission peak with increasing excitation fluence.

The above observations are replicated in a second sample (Fig.4D), where we show the normalized PL emission spectra for a negatively detuned vdW metasurface ($\delta$ = -55 meV). When comparing the relative spectral position of the two emission peaks (Fig.4E), we find an overall shift to lower energies for both polariton branches. Because stronger interaction is expected to shift the polariton to higher energies, this apparent redshift is driven by the peculiar response of the exciton resonance in WS$_2$ (*51*), as also observed for the reference heterostructure in Fig.S9. However, the overall relative blueshift of the lower polariton branch, and redshift of the upper one, confirms that the main process in our metasurfaces is related to the phase state filling mechanism (*47*), where, in analogy with the Pauli exclusion principle, the oscillator strength of the optical transition is reduced due to the lower number of excited states available. Notably, the PL intensity does not exhibit a saturation even under maximum fluence from our excitation source (Fig.4F). This lack of saturation indicates a saturation of the strong coupling regime and transition to a lasing regime (*52*) as previously shown only with theoretical models (*53*). To confirm this effect, we plot in Fig.4G the values of the linewidth for both peaks, extracted from the data presented in Fig.4D. We observe an opposite behavior between the two peaks, with the expected broadening of the exciton under higher densities, owing to the predominance of non-radiative effects. However, for the polariton peak, we observe a reduction in linewidth, consistent with the onset of an amplified spontaneous emission rate (*54*).

In particular, the observed nonlinear polaritonic behaviour in our vdW-HM appear at excitation fluences in the order of nJ cm$^{-2}$, 3 to 4 orders of magnitude lower than previous works on TMDCs monolayers strongly coupled to optical cavities (*39, 42, 55–57*), indicating extremely large exciton-polariton interactions at low polariton populations. Because the metasurface exciton-polaritons are spatially confined in the single unit cell at length scales of hundreds of nanometers, the huge amount of stored kinetic energy, subject to collisional losses and strongly dependent on the confinement length (*58*), could be the origin of such large nonlinearities. Higher Q factors cavities (Q > $10^3$) and low temperature spectroscopy could unveil the spectral signatures of confined exciton-polaritons, which are likely absent in our current samples due to the lower Q factors designed to achieve exciton-cavity critical coupling (*38*). These results open exciting perspectives for room temperature polaritonics, lasing and multibody physics in 2D materials within ultrathin optical devices and enable the reproducible integration of monolayer materials in optical metasurfaces.

In conclusion, we have introduced the concept of van der Waals (vdW) heterostructure metasurfaces, merging the design flexibility of optical metasurfaces with the deterministic assembly of ultrathin vdW material heterostructures. Utilizing excitons in a WS$_2$ monolayer embedded within a nanostructured hBN-based vdW heterostructure, we demonstrated strong light-matter coupling at room temperature. Remarkably, this was achieved without needing an external optical cavity, but by carefully designing optical metasurfaces to generate intrinsic optical resonances within the stacked vdW materials. The characteristic exciton-polariton physics of the embedded $WS_2$ monolayer observed in both absorption and emission showcase extremely strong polariton interactions, more than 3 orders of magnitude larger than in other 2D-cavity systems. Our approach seamlessly integrates 2D materials with flat optics and can be readily applied to a wide range of vdW heterostructures, including those based on TMDCs hetero-bilayers, black phosphorus, and other vdW semiconductors. Following the same principles, our rational vdW-HM design methodology can be extended to other research fields involving more exotic vdW materials, from superconducting to topological, opening new avenues for the investigation of light-matter interactions in complex layered quantum materials. Overall, vdW metasurfaces pave the way for advancements in both nanooptics and condensed matter physics, fostering the development of next generation nanophotonic devices with unparalleled structural control and tailored functionalities for applications in optical communications, quantum technologies, sensors, and beyond.

**Acknowledgments:** Views and opinions expressed are however those of the author(s) only and do not necessarily reflect those of the European Union or the European Research Council Executive Agency. Neither the European Union nor the granting authority can be held responsible for them.

**Funding:**

DFG, German Research Foundation EXC 2089/1 – 390776260

DFG, German Research Foundation MA 4699/7-1

DFG, German Research Foundation TI 1063/1

European Union METANEXT, 101078018

Bavarian program Solar Energies Go Hybrid (SolTech)

Center for NanoScience (CeNS)

Alexander von Humboldt Foundation

Lee-Lucas Chair in Physics

**Author contributions:**

Conceptualization: LS, LK, AT

Methodology: LS, JB, LK

Investigation: LS, JB, LL

Visualization: LS, JB, AT

Funding acquisition: LS, LL, AH, SAM, AT

Project administration: AT, SAM

Supervision: AT, SAM, LdSM, AH

Writing – original draft: LS

Writing – review & editing: LS, JB, AT

**Competing interests:** Authors declare that they have no competing interests.

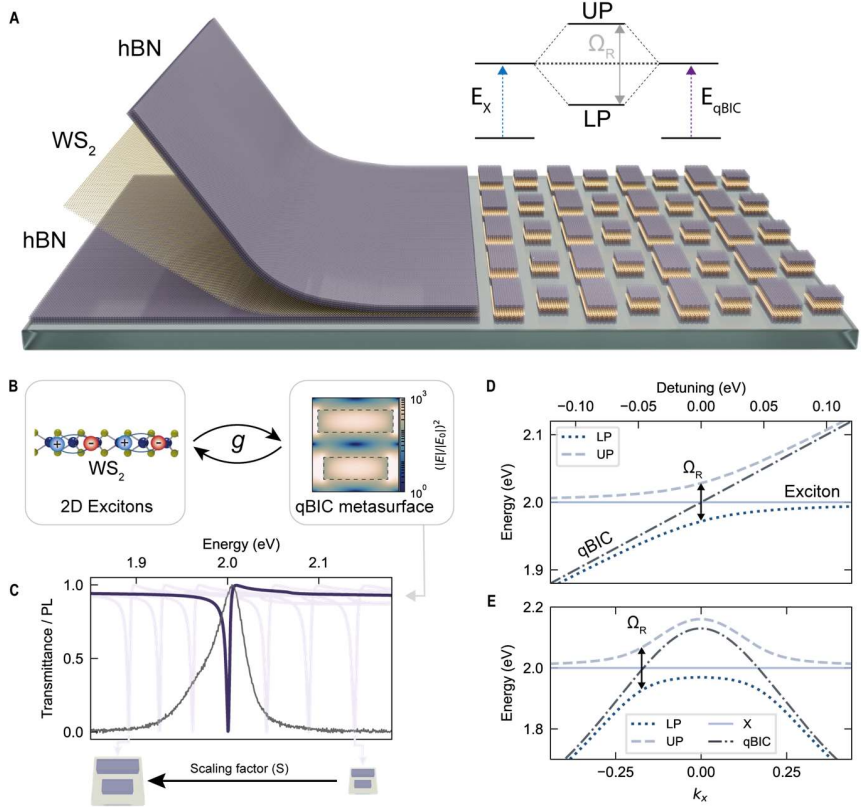

**Fig. 1. Van der Waals heterostructure metasurfaces** (**A**) Illustration of a van der Waals heterostructure metasurface, composed of a $WS_2$ semiconductor monolayer encapsulated between two thin hBN layers. The monolithic metasurface design is patterned into the heterostructure via top-down nanofabrication processes. The asymmetric unit cell design, composed of two nanorods with different length, gives rise to symmetry-protected quasi-bound states in the continuum (qBIC) resonances, which strongly confine light and resonantly interact with the 2D $WS_2$ excitons. Top right: Schematic of the strong coupling regime between excitons ($E_X$) and qBIC resonances ($E_{qBIC}$), with the formation of the upper (UP) and lower (LP) exciton-polariton branches separated by the Rabi energy ($\Omega_R$). (**B**) Strong coupling between the exciton in the $WS_2$ semiconductor (left panel) and the photonic qBIC resonance (right panel) can be reached by maximizing the coupling strength ($g$). Numerical FDTD simulations of the electromagnetic field enhancement $(|E|/|E_0|)^2$ for an hBN metasurface at the qBIC resonance show strong fields inside the resonators, ideally suited for boosting light-matter coupling. (**C**) Numerical simulations of the transmission of an hBN metasurface ($\Delta L = 50$ nm) demonstrate effective spectral tuning of the qBIC resonance via a lateral scaling factor (S) applied to the unit cell. The PL spectrum of the room temperature emission for a monolayer $WS_2$ encapsulated in hBN is shown in black. (**D**) Analytical calculation of the dispersion following a coupled harmonic oscillator model (see Methods) and formation of the LP and UP branches in the strong coupling regime, as a function of the detuning between cavity (qBIC) and emitter (excitons, X). (**E**) Momentum dispersion of the qBIC-exciton system as a function of the wavevector $k_x$, parallel to the nanorod long axis. The solid line and the dot-dashed line represent the exciton and qBIC modes, respectively. The dashed lines are the polariton branches. When the qBIC mode is positively detuned from the exciton, the negative angular dispersion makes the Rabi splitting appear at larger angles, where the two resonances cross.

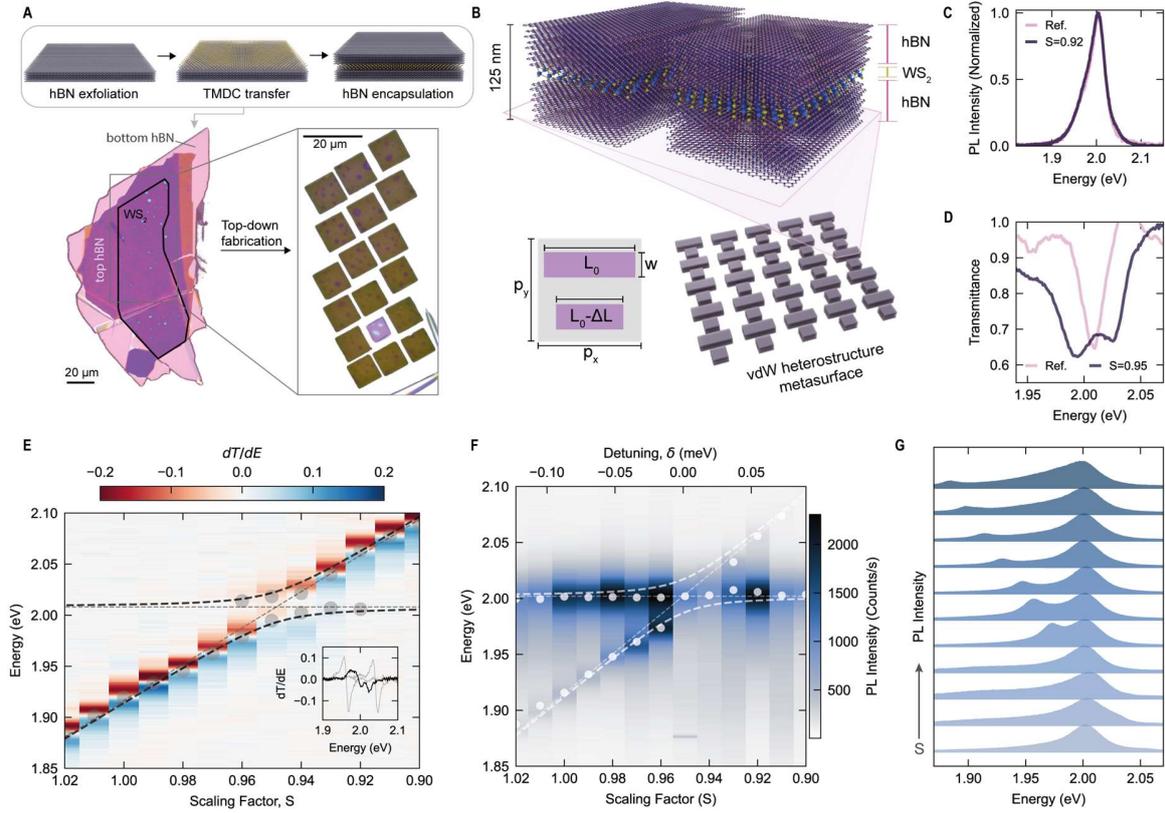

**Fig. 2. Room temperature strong coupling in hBN/WS$_2$/hBN vdW heterostructure metasurfaces.** (**A**) Optical image of the vdW heterostructure. The WS$_2$ monolayer (black outline) is encapsulated between a bottom hBN layer (light pink area) and a top hBN (purple area) using a multi-step heterostructure fabrication process (top panel). Right panel: Optical image of the final vdW heterostructure metasurface sample after lithography-based nanofabrication. The bright square is the unstructured reference heterostructure sample. (**B**) Schematic illustration and side view of a single qBIC metasurface unit cell, composed of hBN/WS$_2$/hBN layers with a total height of 125 nm. Inset: Geometry of the metasurface unit cell, where $p_x = p_y$ is the lattice periodicity, $w$ is the nanorod width, $L_0$ is the base rod length, and $\Delta L$ is the asymmetry factor. (**C**) Photoluminescence (PL) spectra of the reference vdW heterostructure and a positively detuned vdW HM showing negligible changes in the PL emission spectra confirming the preservation of monolayer quality after the fabrication step. (**D**) Transmittance of the reference WS$_2$ heterostructure sample, exhibiting the excitonic peak at 2 eV (multiplied 15x), and that of the resonant qBIC metasurface, exhibiting a clear splitting at the corresponding exciton energy. (**E**) Derivative of the transmittance (dT/dE) for the full set of optical metasurfaces for the sample in Fig.2A, fitted with the coupled harmonic oscillator model described in Methods. Inset: selected derivative trace for vdW HM at close to zero detuning (in black) and of negatively and positively detuned metasurfaces as comparison (in grey). (**F**) PL emission as a function of the scaling factor ($S$) and relative qBIC-exciton detuning ($\delta$). The white dots represent the PL peak maxima, and the data is fitted from the same coupled harmonic oscillator as in Fig.2E. (**G**) Normalized WS$_2$ PL spectra from different vdW metasurfaces with increasing scaling factor (bottom to top).

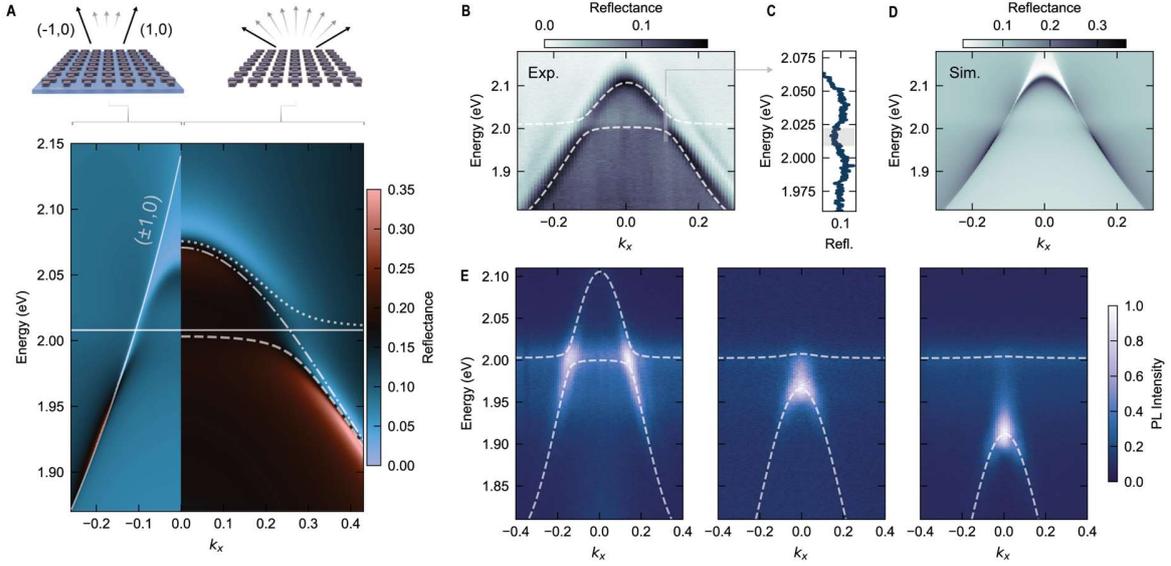

**Fig. 3. Angle-resolved spectroscopy of vdW heterostructure metasurfaces. (A)** Angular dispersion of a vdW HM on a substrate and in vacuum. In the substrate case, light is mainly redirected into the diffraction modes (black arrows in top panel). In vacuum, the grating modes appear at higher angles, revealing the full exciton-polariton dispersion. Bottom panel: RCWA simulations of a vdW heterostructure on a glass substrate (negative wavevectors) and in vacuum (positive wavevectors). The dispersion in vacuum is fitted using the model described in Methods, with the resulting polariton branches depicted as dashed white lines. The solid, flat line represents the $WS_2$ exciton resonance. **(B)** Experimental back focal plane imaging of the reflectance from a vdW metasurface with a detuning $\delta$ = 96 meV. The dispersion is overlayed with the calculated exciton polariton branches energies (dashed white lines, see Methods for details). **(C)** Reflectance spectrum extracted from Fig.3B at $k_x$ = 0.09, exhibiting the splitting at the corresponding exciton energy (shaded grey area) **(D)** RCWA numerical simulations of the reflectance from the vdW heterostructure, showing good agreement with the experimental results, and underscoring the prominent effect of the diffraction modes. **(E)** Angular dispersion of the $WS_2$ PL emission from different vdW-HM with different detuning (96 meV, -55 meV and – 94 meV, from left to right). For negatively detuned cavities, the maximum of the PL emission is observed at $k_x$=0, corresponding to the qBIC-coupled lower polariton branch.

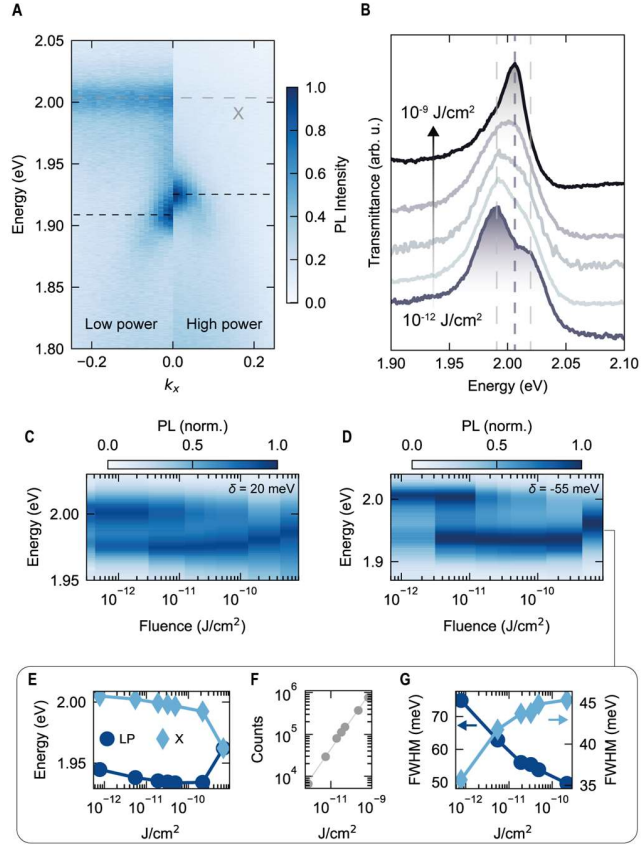

**Fig. 4. Nonlinear exciton-polaritons and saturation of the strong coupling regime (A)** Comparison of the PL angular emission for a vdW-HM ($\delta$ = -94 meV) at low ($10^{-13}$ J cm$^{-2}$) and high ($10^{-10}$ J cm$^{-2}$) excitation fluences. The grey dashed line represents the exciton energy, while the black dashed lines mark the peak position of the LPB emission. **(B)** Transmittance for a vdW-HM ($\delta$ = 20 meV) excited using a supercontinuum laser with a repetition rate of 2.05 kHz. Similarly to the emission, we observe, even in absorption, a transition to weak coupling at fluences above $10^9$ J cm$^{-2}$. **(C)** Normalized PL emission for a metasurface, $\delta$ = 20 meV, as a function of the excitation fluence. The exciton peak in the emission doublet is prominent at low power, but it gradually loses its weight, leaving the lower polariton branch to dominate the PL emission until fluences approaching $10^9$ J cm$^{-2}$, where we observe a collapse to a single optical transition. **(D)** Normalized PL emission of a vdW metasurface on a second sample ($\delta$ = -55 meV) under increasing fluences. We observe the same saturation of the strong coupling regime when approaching $10^{-9}$ J cm$^{-2}$ excitation fluences. **(E)** Peak maxima positions for the exciton (X) and LP peaks for increasing excitation fluence. Both peaks exhibit a redshift with increasing power, eventually leading to the collapse of both peaks into a single emission channel. **(F)** Integrated PL intensity as a function of the excitation fluence, exhibiting no saturation up to values approaching $10^9$ J cm$^{-2}$. **(G)** Full width at half-maximum (FWHM) values extracted for the exciton peak and LPB peak as a function of the excitation fluence. The exciton exhibits the expected broadening due to the increased non-radiative processes, while on the contrary, for the LP, we observe a reduction of the linewidth.

# Supplementary Materials for

**Van der Waals heterostructure metasurfaces: atomic-layer assembly of ultrathin optical cavities**

Luca Sortino *et al.*

Corresponding authors: luca.sortino@physik.uni-muenchen.de, Andreas.Tittl@phyisik-uni.muenchen.de

**Materials and Methods**

Fabrication and patterning of the van der Waals heterostructures

The van der Waals heterostructures were fabricated by mechanical exfoliation and a combination of dry transfer techniques from commercially available single crystals (HQ Graphene). Initially, hexagonal boron nitride (hBN) was exfoliated onto commercial silicon wafers. Flakes with thickness ranging from 60 to 80 nm were selected using optical microscopy and a Stylus profilometer (Bruker Dektak) to achieve a total height of approximately 120 nm to 160 nm of the final heterostructure. A first hBN flake was transferred onto a fused silica ($SiO_2$) substrate using a polydimethylsiloxane (PDMS) and poly(bisphenol A carbonate) (PC) stamp. Subsequently, tungsten disulfide ($WS_2$) was mechanically exfoliated onto a separate PDMS stamp. Suitable monolayers were detected via optical microscopy and then transferred on top of the first hBN flake. The second hBN flake was then picked up using a PDMS and PC stamp and transferred on top of the $WS_2$ monolayer and base hBN-flake complex, thus completing the van der Waals heterostructure.

Stacks were further processed and patterned following the methods described by Kühner et al. (*35*). As a metasurface unit cell design, a double rod system with length asymmetry was chosen. At a scaling factor $S = 1$, the pitch size was set to $p_x = p_y = 410$ nm, and the resonators were parameterized by length $L_0 = 360$ nm, width of $w = 100$ nm and asymmetry factor $\Delta L = 75$ nm. A double-layer of poly(methyl methacrylate) (PMMA, Allresist, 80 nm of 950k on top of 100 nm of 495k) was spin-coated onto the substrate, and each PMMA layer baked at 170°C for 3 minutes. Electron beam lithography was then employed to define the drafted metasurface pattern in the PMMA layer on top of the monolayer region. After development, 2 nm titanium (Ti) as adhesion promoter and 60 nm gold (Au) were deposited using electron beam evaporation. Bathing the sample over 8 h in a resist remover solution (Microposit remover 1165) allowed for a clean lift-off process and creates the hard-mask pattern of choice. The unprotected regions of the heterostructure were etched away using inductively coupled reactive ion etching (ICP-RIE) with sulfur hexafluoride ($SF_6$) and argon (Ar) gases under 6.0 mTorr pressure with 300 W HF and 150 W ICP power. The remaining gold hard-mask was then removed inside a potassium-based Au etchant.

Optical spectroscopy

The optical spectroscopy of vdW heterostructure metasurfaces was carried out in a homebuilt optical setup working in transmission and reflection geometry, as shown in Fig.S5. For reflection measurements, the collimated emission of a tungsten-halogen source (Thorlabs SLS201L) is sent through a 50:50 beamsplitter and into the aperture of a 0.95 numerical aperture (NA) 60x objective. The reflected white light signal is collected through the same objective and focused on the slit of a spectrometer with 75 mm focal length (Princeton Instruments) for spectroscopic analysis.

For angle-resolved spectroscopy, the addition of a Bertrand lens (BL in Fig.S5) in the collection path, focusing at the back focal plane of the objective and in a 4-*f* system with the lens in front of the spectrometer, allows to project on the slits the angle resolved signal collected form the sample. To achieve spectral dispersion, the sample is mounted on a rotating stage to align its axis with the spectrometer slits. By utilizing the vertical slits as a spatial filter, the reflected signal is projected onto the grating, dispersing the spectral components onto the spectrometer CCD camera.

For photoluminescence measurements, the excitation is performed with the output of a frequency doubled, 80 MHz repetition rate, 150-fs pulsed Ti:Sapphire laser (Coherent Chameleon Ultra II) tuned to 405 nm. The laser is aligned in a transmission geometry and focused on the sample through a 0.25 NA 10x objective. The resulting focused laser spot size resulted in 8.15 µm in diameter, which is large enough to efficiently excite the optical metasurfaces (approximately 10 µm x 10 µm in size). The transmitted PL emission is then directed to the detector via the same setup described for reflection measurements. To obtain a spectral dispersion, we aligned the long axis of the metasurface nanorods to the spectrometer slit and used it as a spatial filter. The dispersion with the spectrometer optical grating allows the retrieval of the angular dispersion of the light emitted by the metasurface. For the measurements in transmission, the sample is either excited with a halogen lamp, or with the output of a supercontinuum ~35 ps pulsed laser (NKT photonics), at a repetition rate of 2.05 kHz.

Numerical methods

Numerical simulations are carried out via a commercial software package (Lumerical Ansys) employing a rigorous coupled-wave analysis (RCWA) method for the angular dispersion of the metasurfaces and a finite-difference time-domain (FDTD) package for the calculation of the electromagnetic fields. The monolayer is simulated with a mesh of 0.1 nm, ensuring a high number of mesh cells, with a nominal thickness of 1 nm.

Coupled harmonic oscillator model

The qBIC mode dispersion at low angles can be represented as the diffractive coupling of two symmetry-protected counter-propagating in-plane modes (46). The qBIC spectral position as a function of the emission angle $\theta$ is modelled as (59):

$$E_{\text{qBIC}}(\theta) = E_{\text{qBIC}}(\theta = 0) + U - \sqrt{U^2 + v^2 k(\theta)^2} \quad (1)$$

where, $E_{\text{qBIC}}(\theta = 0)$ is the value at zero emission angle, $U$ is the off-diagonal coupling term, and $v$ represents the group velocity of the BIC mode at high angles, $k(\theta) = (2\pi/a)\sin\theta$ is the in-plane wavevector and $a$ denotes the lattice periodicity. The Hamiltonian of the coupled harmonic oscillator model describing the coupling between the qBIC mode ($E_{qBIC}$) and the 2D excitons ($E_X$), takes the form of:

$$H = \begin{pmatrix} E_{qBIC} - i\gamma_{qBIC} & g \\ g & E_X - i\gamma_X \end{pmatrix} \quad (2)$$

where $\gamma_{qBIC}$ and $\gamma_X$ are the damping rates of the qBIC mode and exciton, respectively, and $g$ is the coupling strength. Solutions of the model for different exciton-qBIC detuning values are shown in Fig.S1. A value of $\gamma_X$ = 14 meV is extracted from the fit of the optical transmission from the encapsulated sample and we extract $\gamma_{qBIC}$ = 6.5 meV from numerical simulations and values of approximately 15 meV from experimental data fitting (Fig.S3).

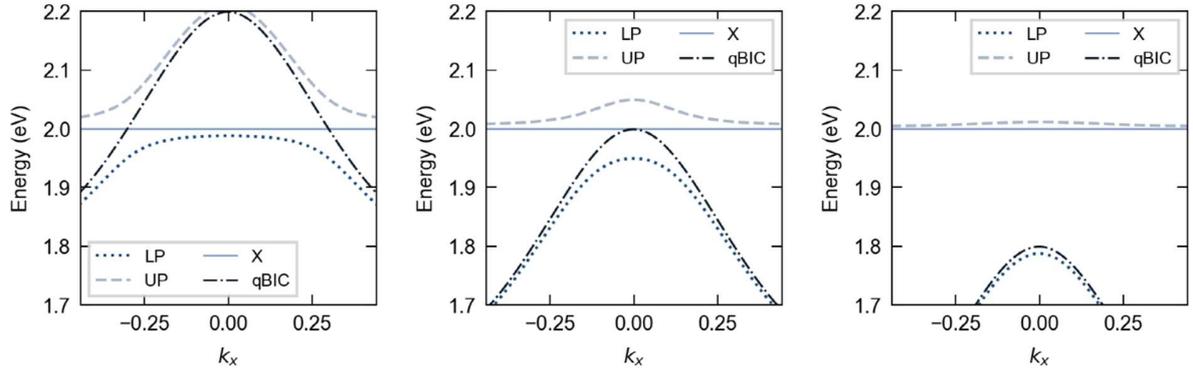

**Fig. S1.** Angular dispersion of the qBIC-exciton coupled system, calculated with the model discussed in Equations 1-2 in Methods. The exciton and qBIC are showed in the case of positive detuning $\delta = E_{qBIC} - E_X$ (left panel), zero detuning (central panel) and for negative detuning (right panel).

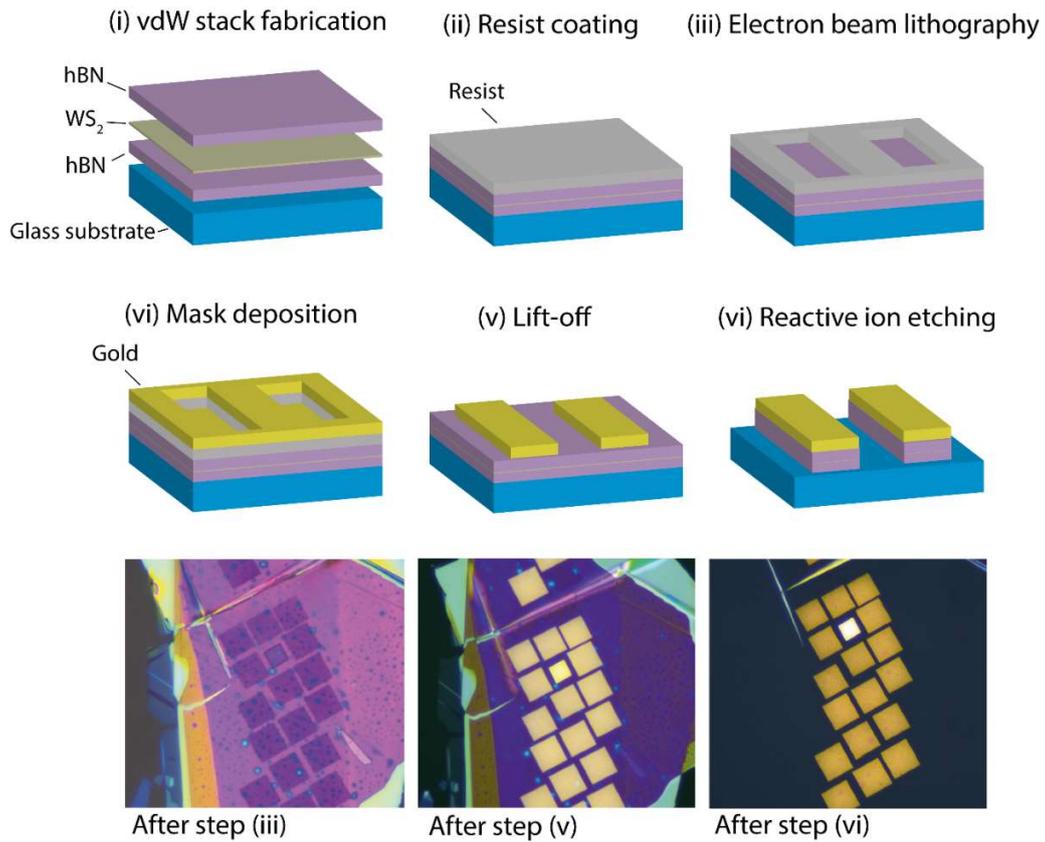

**Fig. S2.** (i-vi) Schematics of the fabrication method. **(i)** Fabrication of the van der Waals stack on glass substrate. **(ii)** Deposition via spin coating of the resist. **(iii)** Imprinting of the design in the resist via electron beam lithography and subsequent development. **(iv)** deposition of the thin film gold etching mask. **(v)** Removal of the excess gold leaving via lift-off. **(vi)** Reactive ion etching and removal of excess materials, leaving only the metasurface with the etching mask, later removed via a gold etchant (step not shown). Bottom row: optical microscope images of the sample after step (iii), (v) and (vi), from left to right.

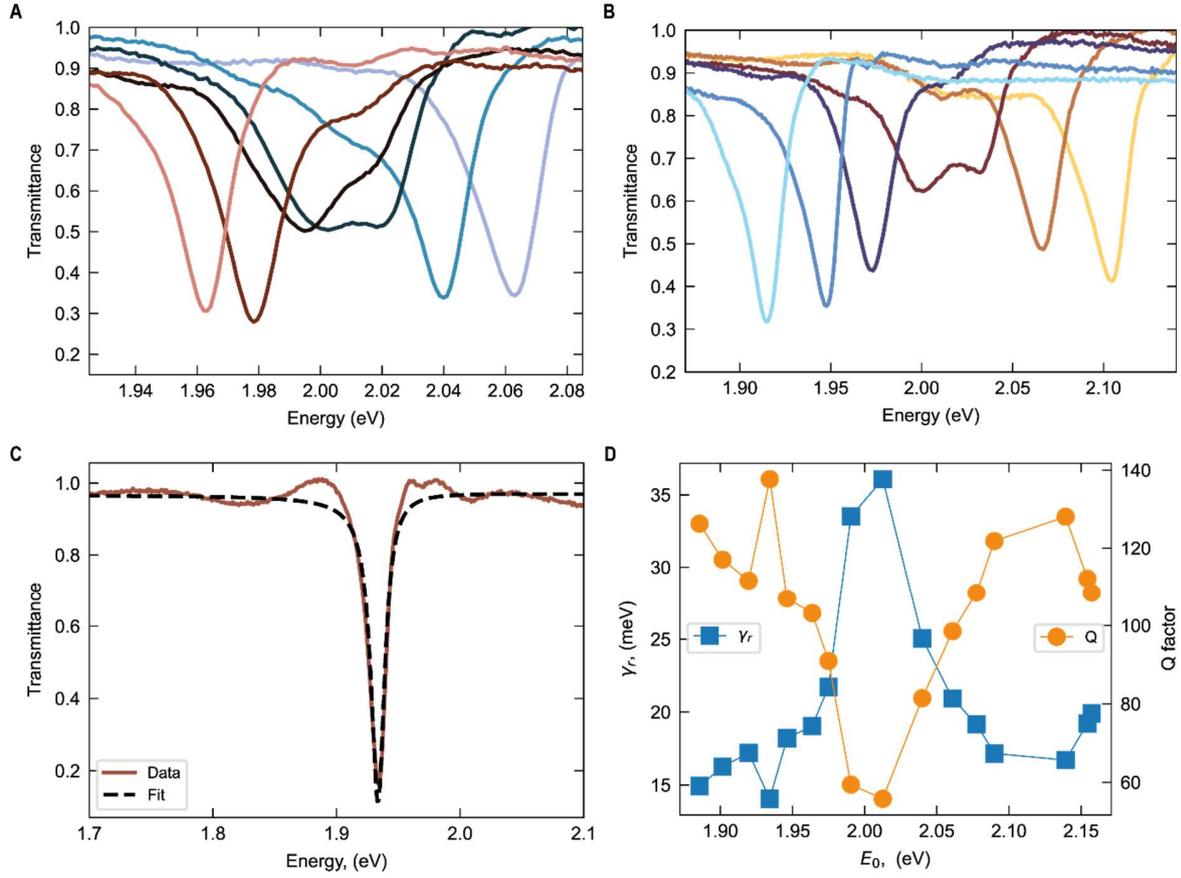

**Fig. S3. (A)** Normalized transmittance spectra for sample 1 shown in the main text. **(B)** Normalized transmittance spectra for the sample shown in Fig.S3. **(C)** Example of the fitting procedure of the optical transmission of a vdW heterostructure metasurface (dashed black line), following the Fano formula in Ref.(*6*) . **(D)** Values of the qBIC damping rate ($\gamma_r$, blue squares) and quality factor (Q, orange circles), defined as $Q=E_0/\gamma_r$, where $E_0$ is the qBIC resonance energy extracted from the Fano fit. We observe Q factors above 100 outside of the $WS_2$ exciton resonance, the fit formula yields larger linewidths and lower Q factors around the exciton energy, due to the splitting and line broadening when in resonance.

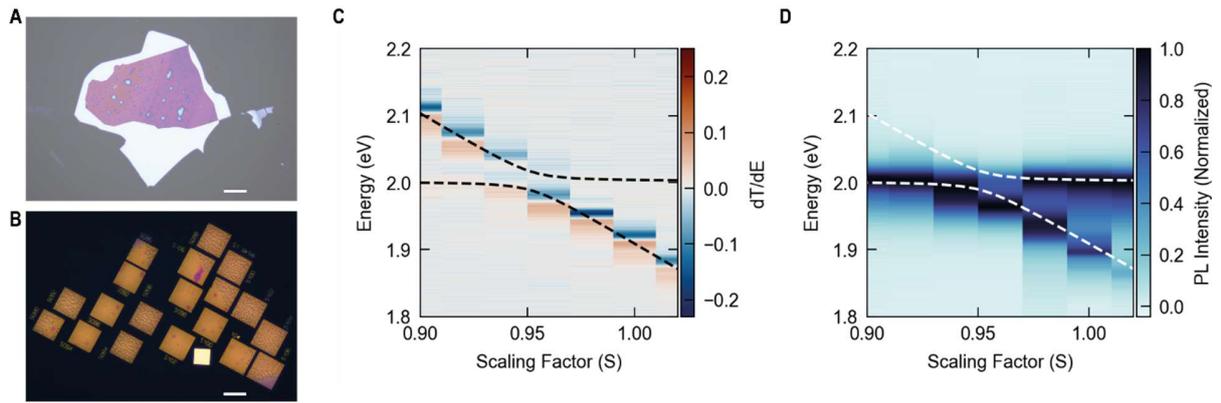

**Fig. S4. (A)** Optical image of a second sample used in this study, before the nanofabrication of the optical metasurface. Scale bar: 100 µm. **(B)** Optical image of the sample after the last etching step. Scale bar: 20 µm. **(C)** Derivative of the transmission of the sample for different scaling factors, fitted as discussed in Fig.2 in the main text (dashed black lines). **(D)** Normalized photoluminescence emission as a function of the scaling factor, exhibiting the red shifted polariton peak following the qBIC mode dispersion.

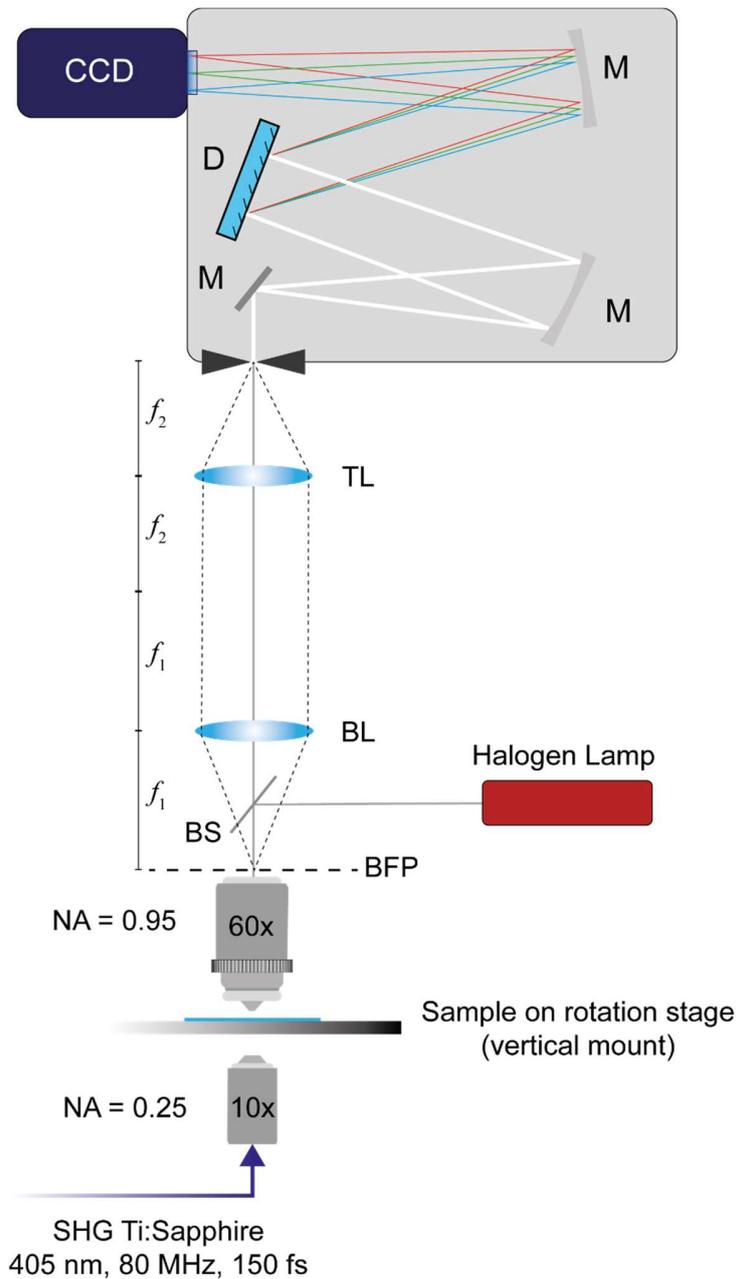

**Fig. S5.** Schematics of the back focal plane spectroscopy setup. The sample is mounted vertically in respect to the optical table, and on a rotation stage, to align the metasurface to the vertical spectrometer slit. The sample is imaged by sending a halogen lamp emission through a beamsplitter (BS), and in a high numerical aperture (NA) objective. The reflection is collected in a 4-$f$ setup to image the back focal plane (BFP) of the same objective. For the study on PL emission, the sample is excited in a transmission setup with the emission of a frequency doubled Ti:Sapphire fs-pulsed laser. Removing the Bertrand lens (BL) allows to focus the real space image of the sample via the tube lens (TL) in front of the spectrometer, for conventional μ-spectroscopy analysis.

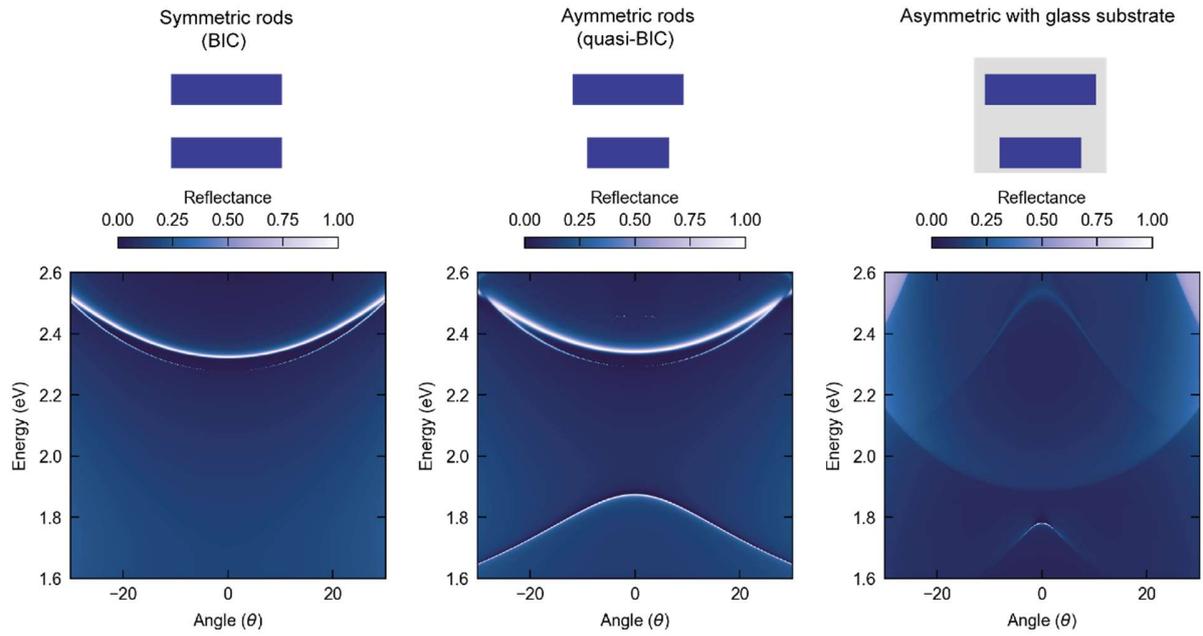

**Fig. S6.** Evolution of the RCWA numerically calculated angular dispersion from a symmetric hBN metasurface (left panel), in vacuum, and adding the asymmetry factor, thus creating the qBIC state (central panel) with the observed negative dispersion at higher angles. In the right panel, the same qBIC metasurface, but now with the additional $SiO_2$ glass substrate, highlighting the impact of the grating modes on the angular dispersion of the qBIC resonance.

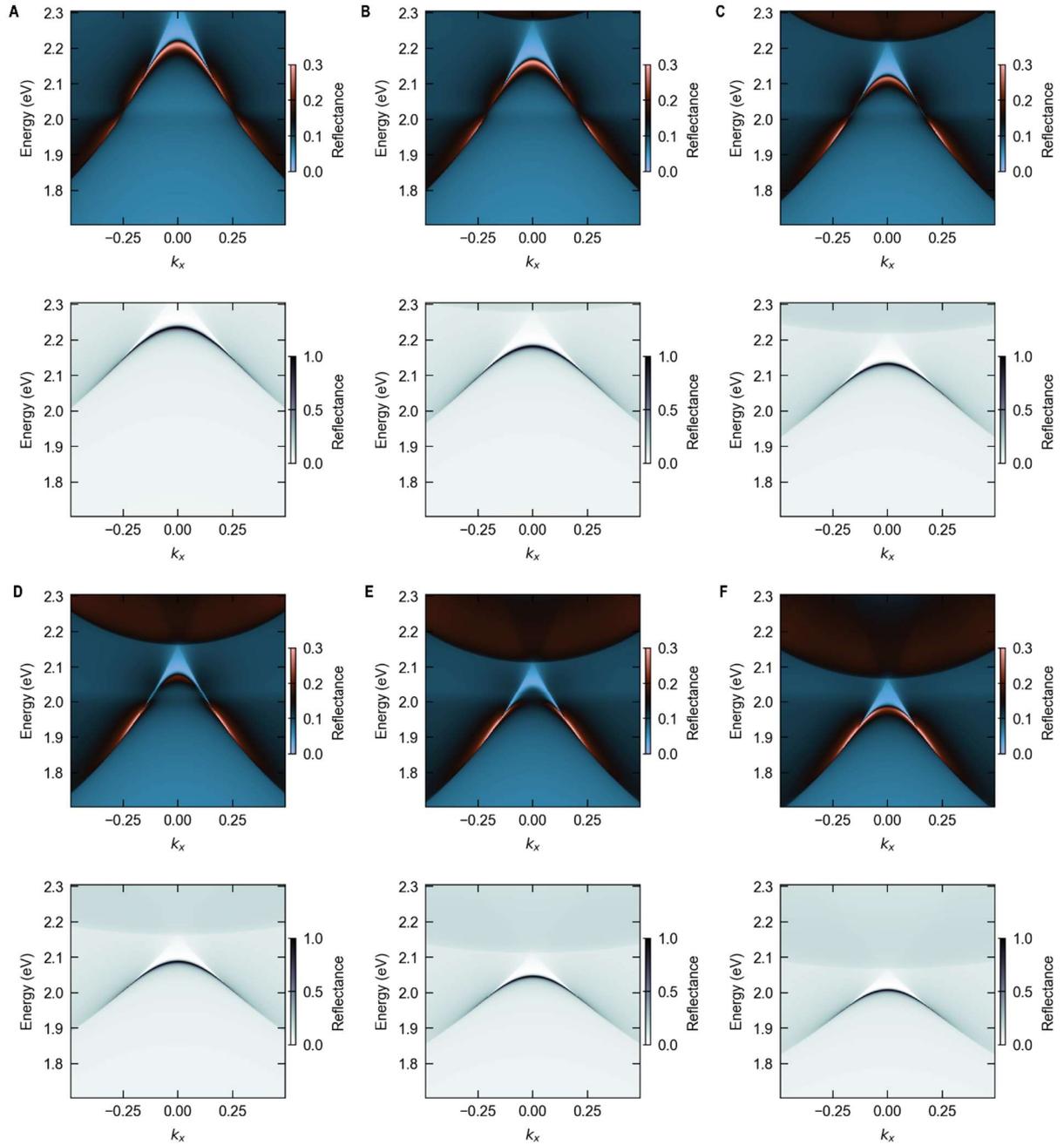

**Fig. S7.** RCWA numerical simulations for hBN metasurfaces (125 nm height), on glass substrate and with 1 nm of $WS_2$ embedded at half-height of the hBN resonator, are shown in the top panels with the blue/red colorscale. The bottom panels (grey colorscale) show the same hBN metasurface but without the 2D monolayer (bottom panels). The qBIC cavity mode is tuned across the exciton resonance energy, by gradually increasing the scaling factor (A to F).

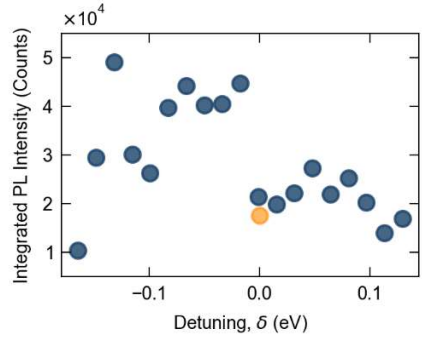

**Fig. S8.** Integrated PL intensity for the sample presented in Fig.2 in the main text. The orange dot represents the value of the reference hBN encapsulated $WS_2$. For positively detuned cavities, we observe negligible PL enhancement. Instead, negatively detuned cavities exhibit a higher PL emission, related to the increased light-matter coupling and polariton PL emission observed as an additional peak in the PL emission spectra.

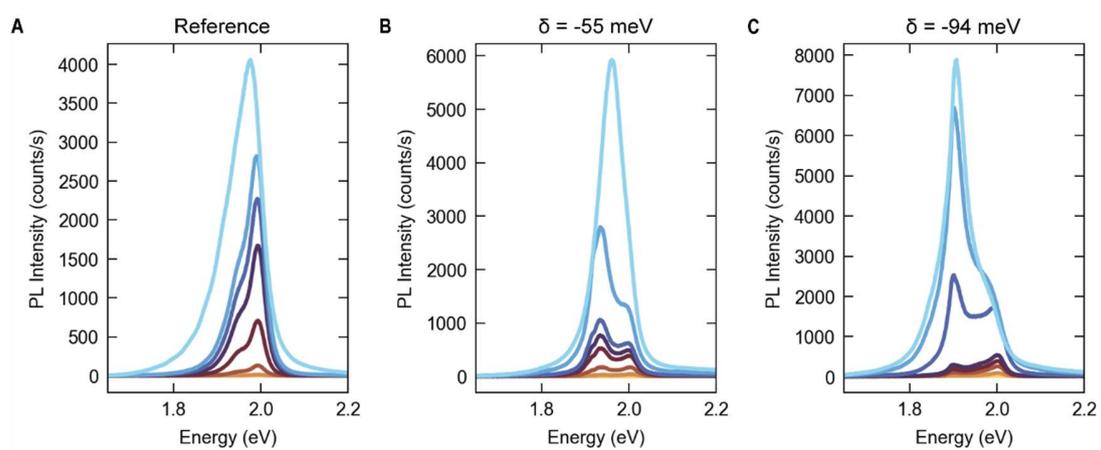

**Fig. S9. (A-C)** Photoluminescence (PL) spectra under increasing excitation fluences, color coded as in Fig.4D in the main text, illustrated for (A) the reference sample of an hBN encapsulated $WS_2$ monolayer, in (B) a vdW metasurface with $\delta$ = -55 meV and in (C) with $\delta$ = -94 meV.